\RequirePackage{fix-cm}
\documentclass[smallextended]{svjour3}       
\smartqed  
\usepackage{graphicx}
\usepackage{amssymb,amsmath}
\usepackage{bm,enumerate}
\usepackage{textcomp}
\usepackage[colorlinks=true,citecolor=blue,urlcolor=blue]{hyperref}
\usepackage[utf8x]{inputenc}
\usepackage{cite}

\begin{document}

\title{ Quantum Coherence and non-Markovianity in a Noisy Quantum Tunneling Problem}

\titlerunning{}

\author{Nisreen Mohammed Mahdi$^{1}$, Arzu Kurt$^{1}$, and Ferdi Alt{\i}nta\c{s}$^{1}$}

\institute{Arzu~Kurt\\
\email{arzukurt@ibu.edu.tr}\\
\\
$^1$ Department of Physics, Bolu Abant Izzet Baysal University, Bolu 14030, T\"{u}rkiye.
}

\date{Received: date / Accepted: date}

\maketitle

\begin{abstract}

We investigate the coherence and non-Markovianity of a quantum tunnelling system whose barrier is fluctuated by a telegraph noise, and its energy gap is modulated by a Gaussian noise. With the help of averaging method, the system dynamics is analytically derived, and the analytical expression for coherence measure and non-Markovianity for the very limited parameter regimes for both initially coherent and non-coherent states are obtained. We observe non-Markovian dynamics in a situation where Kubo number is high. It is also found that there is no strong relation between coherence of the system and non-Markovianity dynamics except a region in which these two tend to change their behavior at the intermediate noise color for two initial states.


\end{abstract}
\section{Introduction}

One of the quasi-essential problems of quantum mechanics is the dynamics of a particle moving in a double-well potential~\cite{CohenTannoudji} which can be used a model system to introduce quantum concepts such as tunneling through a barrier which plays an important role in determining the structure of matter and is exploited to develop many nanotechnological devices~\cite{Trixler2013,Taylor2004,Ionescu2011} and two-level system (TLS) approximation which is utilised widely in many fields from magnetic resonance to quantum information theory to model foundational and practical problems. While analytical expressions for the dynamics of a closed TLS are easy to obtain, the unavoidable interaction between the environment and the TLS, which is responsible for the dissipation and decoherence, complicates the problem and there is still no analytical solution for the dynamics of a TLS coupled to a quantum mechanical thermal bath~\cite{Leggett1987,Iwans2000,Weiss1993}. One of the widely used approaches to deal with this difficulty is to model the quantum mechanical environment as a 
classical noise that perturbs the level energies of the TLS~\cite{Cukier1997,Grifoni1998,Grifoni1996,Grifoni1997,Goychuk1995,Goychuk21995,Petrov1996,Petrov21996,Goychuk1997}. The environmental perturbations or external driving could also affect the tunneling rate or the potential barrier height~\cite{Grifoni1996,Grifoni1997,Goychuk1995,Goychuk21995,Petrov1996,Petrov21996,Goychuk1997}. The tunneling problem with time modulated parameters has been known to occur in several physical situations, for example long-range electron transfer reactions~\cite{Makri1996,Daizadeh1997}, some problems in semiconductor physics~\cite{Weisbuch1991}, polarization of a multilayer structure~\cite{Zhu2004,Daninthe2006} and the scattering problem~\cite{Buttiker1982,Burmeister1998}.

Quantum coherence is one of the unique features of quantum mechanics, indicating the ability of a state to exhibit quantum interference effects and enabling quantum features such as quantum entanglement, non-locality, discord, and steering. It plays an important role in quantum information and computation protocols~\cite{Nielsen2000}, and also in emerging areas, such as quantum thermodynamics~\cite{Lostaglio2015}, biology~\cite{Plenio2008}, and metrology~\cite{Dobrzanski2014}. Only recently has a rigorous theory for measuring quantum coherence developed with the aim of adopting coherence as a physical resource~\cite{Baumgratz2014,Yu2016,Yu22016}. After the development, its intimate connection to the entanglement and quantum correlations have been discussed in a number of settings~\cite{Streltov2015,Yao2015,Xi2015,Ma2016,Radhakrishnan2016}. Furthermore, coherence freezing and distillation have been investigated in~\cite{Winter2016,Bromley2015,Yu2016,Yu22016,Chitambar2016}. On the other hand, non-Markovianity is another quantum mechanical effect, reflecting the back-action of the environment on the quantum systems. Exploring non-Markovian features can be important in the fabrication of nanoscale machines~\cite{Ribeiro2015,Abiuso2019,Pezzutto2019,Katz2016,Ptaszynski2022,Zhang2014}. Specifically, in the context of quantum thermodynamics, non-Markovianity is used as an additional fuel to supply a quantum thermal machine~\cite{Ptaszynski2022,Zhang2014}. Along similar lines, a measure for the degree of non-Markovianity, which accounts for non-zero values for the information flow back to the system, has recently been reported~\cite{Breuer2009,Laine2010}. One of the interesting questions is whether there exists any relation between the dynamics of coherence and the non-Markovianity of the system dynamics~\cite{Addis2014,Cakmak2017,Man2018,Bhattacharya2018,Chanda2016, He2017, Mirafzali2019, Radhakrishnan2019,Kurt2020,Liu2018}. Addis et. al. have shown that the stationary coherence is maximized when the dynamics is non-Markovian for pure dephasing model~\cite{Addis2014}, while Bhattacharya et. al.~\cite{Bhattacharya2018} have observed that non-Markovianity might be relevant for coherence as a resource for quantum tasks. A number of authors have proposed possible measures of non-Markovianity based on the various coherence measures~\cite{Chanda2016,He2017,Mirafzali2019}. Radhakrishnan et al. have observed that coherence oscillates with a decaying envelop or decays exponentially in the
non-Markovian and Markovian regimes, respectively~\cite{Radhakrishnan2019}.

In the current paper, our aim is to study quantum coherence and non-Markovianity of the dynamics of a particle undergoing quantum tunneling in the presence of a stochastic bias and a fluctuating barrier under the two-level approximation. We assume that the fluctuations in the barrier are induced by dichotomous noise, whereas the transition energy of the TLS is perturbed by delta-correlated Gaussian noise, which mimics the coupling of the TLS to a high-temperature reservoir. Recently, tunneling dynamics between two localized energy levels has been investigated in terms of the correlation time of the barrier fluctuations~\cite{Iwans2000}. Resonant damping of the tunneling has been observed due to noise in the system. It is well known that, depending on its properties, external noise can induce non-Markovianity~\cite{Kurt2018}. One of the questions we would like to address here is whether there exist any relations between the non-Markovianity and the coherence. We have analyzed the dependence of the quantum coherence dynamics on the correlation time of the barrier fluctuations, and elucidated its non-Markovian character. We have found that simple analytical expressions for both coherence and non-Markovianity measures could be derived for a restricted range of noise parameters, and they display prominent features at certain system parameters. 

The outline of the paper is organized as follows. In Sec.~\ref{sec:model}, we present the model and its exact analytical solution. Sec.~\ref{sec:measures} introduces the quantum coherence and non-Markovianity measures. In Sec.~\ref{sec:results}, the main findings of the current paper is included and we briefly conclude the main results in Sec.~\ref{sec:conc.}.

\section{The model and its solution}
\label{sec:model}
The problem under consideration is tunneling through a fluctuating barrier in the presence of dissipation and the two-level approximation. It is isomorphic to the problem of a spin-$1/2$ under transverse (along the $x$ axis) and longitudinal (along the $z$ axis) magnetic fields with noisy components~\cite{Iwans2000}. The corresponding Hamiltonian can be given as $(\hbar=1)$~\cite{Iwans2000,Voss2000}
\begin{equation}
\label{eq:ham1}
H = \frac{\epsilon}{2} \,\sigma_ z+ \frac{\epsilon(t)}{2}\,\sigma_z- \frac{\Delta(t)}{2}\,\sigma_x, 
\end{equation}
where $\sigma_i~(i= x,y,z)$ are the standard Pauli operators. Here, $\epsilon$ describes the static energy of the localized potential wells and $\epsilon(t)$ is its noisy component. The noise here is considered to arise from coupling to a high temperature thermal reservoir~\cite{Haken1973,Reineker1982} which is modeled as delta-correlated Gaussian noise (white noise) with zero mean $\langle\epsilon(t) \rangle=0$ and correlation function $\langle\epsilon(t)\epsilon(s)\rangle=2\,\kappa\,\delta(t-s)$; $\kappa$ is the intensity of the noise. $\epsilon(t)$, leading to stochastic energy bias, introduces dissipation in the form of dephasing during the time evolution. The last term in Eq.~(\ref{eq:ham1}) corresponds to tunneling between two energy levels through a fluctuating barrier. We consider the fluctuations to be modulated by random telegraph noise (RTN) $\eta(t)$. Therefore, $\Delta(t)$ is a random function of time that has an explicit form such as $\Delta(t)=\Delta_0+\Delta_1\,\eta(t)$. The noise as identified by the parameter $\eta(t)$ has zero mean ($\langle\eta(t)\rangle=0$) and exponentially decaying correlation function ($\langle\eta(t)\eta(s)\rangle=e^{-2\nu|t-s|}$). RTN, also known as dichotomic noise, describes a signal that jumps between values $+1$ and $-1$ with a rate $\nu$.

The dynamics of the system whose Hamiltonian is given in Eq.~(\ref{eq:ham1}) is described by the Liouville-von Neumann equation:
\begin{equation}
\label{eq:vonNeumann}
\frac{d\,\rho}{dt}=-i\left[H,\rho\right]
\end{equation} where $\rho$ is the density matrix which encodes all information about the system. In order to investigate the effect of the stochastic processes on the system dynamics, the density matrix $\rho$ is averaged independently over the noises $\epsilon(t)$ and $\eta(t)$. We first employ an averaging process over the Gaussian noise $\epsilon(t)$ which results~\cite{Kiely2021}
\begin{eqnarray}
\label{eq:masterEG}
\frac{d\langle\rho\rangle}{dt}=-i\left[H_0,\langle\rho\rangle\right]+2\,\kappa\,(\sigma_z\langle\rho\rangle\sigma_z-\langle\rho\rangle),
\end{eqnarray}
where $\langle\dots\rangle$ indicates the average over the Gaussian noise $\epsilon(t)$ and $H_0=\epsilon\,\sigma_z/2-\Delta(t)\sigma_x/2$ is the Hamiltonian containing free and RTN noise terms.

As a second procedure, we use averaging on the master equation~(\ref{eq:masterEG}) over the RTN noise $\eta(t)$. Using the Bloch vector representation of the density matrix $\rho=(I+\sum_{i=x,y,z}\,P_i(t) \sigma_i)/2$ and exploiting the Loginov-Shapiro theorem~\cite{Shapiro1978}, one can obtain a set of first-order coupled differential equations as functions of the averaged system parameters $\langle\langle P_{i}(t)\rangle\rangle$ and the noise correlators $\langle\langle\eta(t)P_{i}(t)\rangle\rangle$ as
\begin{equation}
\label{eq:genEqu}
    \frac{d\overrightarrow{Y}}{dt}=M\, \overrightarrow{Y}
\end{equation}
where 
\begin{eqnarray*}
\overrightarrow{Y}=\left(\begin{array}{c}
   \langle  \langle P_{x}(t)\rangle\rangle\\
     \langle\langle P_{y}(t)\rangle\rangle\\
    \langle \langle P_{z}(t)\rangle\rangle\\
    \langle \langle\eta(t)P_{x}(t)\rangle\rangle\\
   \langle  \langle\eta(t)P_{y}(t)\rangle\rangle\\
   \langle  \langle\eta(t)P_{z}(t)\rangle\rangle\\
\end{array}\right) \mathrm{and}\,\, M=\left(\begin{array}{cccccc}
     -4\kappa&-\epsilon&0&0&0&0  \\
    \epsilon&-4\kappa&\Delta_0&0&0&\Delta_1  \\
     0&-\Delta_0&0&0&-\Delta_1&0 \\
      0&0&0&-2\nu-4\kappa&-\epsilon&0 \\
      0&0&\Delta_1&\epsilon&-2\nu-4\kappa&\Delta_0 \\
      0&-\Delta_1&0&0&-\Delta_0&-2\nu \\
    \end{array}\right)
\end{eqnarray*}
We would like to remark here that the solution in Eq.~(\ref{eq:genEqu}) is exact and completely describes the dynamics of the system.

\section{Quantum coherence and non-Markovianity}
\label{sec:measures}
A complete theory for the quantification of quantum coherence has been given very recently based on the conditions in analogy to the entanglement theory~\cite{Baumgratz2014,Yu2016,Yu22016}. The main aim in such an attempt is to adopt the quantum coherence as useful resource in the protocols of quantum information and computation. The quantification of coherence is based on a valid metric $\Delta(\rho||\sigma$) that measures the distance between two quantum states $\rho$ and $\sigma$. It can be written as~\cite{Baumgratz2014}:
\begin{equation}
\label{eq:qCoh}
C_\Delta(\rho)=\min_{\sigma\in\,I}\Delta(\rho||\sigma).
\end{equation}
Here $\sigma=\sum_{i=1}^{d}\sigma_{ii}|i\rangle\langle i|$ is the density operator in the space $I$ representing the incoherent quantum states defined in terms of a particular fixed basis $|i\rangle$--in our case, the eigenstates of $\sigma_z$ are chosen as the fixed basis. For a given metric $\Delta(\rho||\sigma)$, $C_\Delta(\rho)$ determines the minimal distance between the coherent state $\rho$ and the set of all incoherent quantum states defined in the subspace $I$. Refs.~\cite{Baumgratz2014,Yu2016,Yu22016} set the necessary conditions to qualify $C_\Delta(\rho)$ as a coherence monotone. It has been demonstrated that not all metrics become a proper coherence measure~\cite{Baumgratz2014,de Oliveira2016, Zhang2015,Yu2016,Yu22016}. On the other hand, $l_1$-matrix norm $\Delta_{l1}(\rho||\sigma)=||\rho-\sigma||_{l_1}=\Sigma_{ij}|\rho_{ij}-\sigma_{ij}|$ and the relative entropy $\Delta_{rel.ent}(\rho||\sigma)=\mathrm{Tr}(\rho\log_2\rho-\rho\log_2\sigma)$ have been shown to satisfy all the requirements for qualifying $C_\Delta(\rho)$ as a coherence measure. Furthermore, the minimum in Eq.~(\ref{eq:qCoh}) for both metrics can be obtained for the diagonal density operator $\sigma\equiv\rho_{diag}=\sum_i\rho_{ii}|i\rangle\langle i|$, where $\rho_{ii}=\langle i|\rho|i\rangle$. Therefore, the quantifiers which are called the $l_1$-norm of coherence--$C_{l_1}(\rho)$-- and the relative entropy of coherence--$C_{rel.ent}(\rho)$-- have the explicit forms~\cite{Baumgratz2014}:
\begin{eqnarray}
C_{l_1}(\rho)&=&\sum_{i\neq j}|\rho_{ij}|\label{eq:cl1},\\
C_{rel.ent}(\rho)&=&S(\rho_{diag})-S(\rho)\label{eq:re},
\end{eqnarray}
where $S(\rho)=-\mathrm{Tr}(\rho\log_2\rho)$ is the von Neumann entropy. It is worth noting that in our calculations we have observed that $C_{l_1}(\rho)$ and $C_{rel.ent.}(\rho)$ give qualitatively the same predictions. Therefore, we will restrict our attention to only $C_{l_1}(\rho)$. Furthermore, the $l_1$ norm of coherence serves a very intuitive quantification of coherence, since it is directly connected to the off-diagonal elements of $\rho$ in a fixed basis ${|i\rangle}$. Recently, some established coherence monotones based on different metrics, such as the (modified) trace norm and fidelity, have been found to be increasing functions of $C_{l_1}(\rho)$~\cite{Zhang2015,Chen2018,Yang2018}. Non-zero $C_{l_1}(\rho)$ is the indication of quantum interference phenomena between two localized potential wells which are represented by the eigenstates of $\sigma_z$. The role of the barrier and its fluctuations in this phenomenon will be investigated for two initial states; one of them is the incoherent state $\rho_1(0)=|1\rangle\langle1|$ where the particle is localized in one of the wells and the other is the maximal coherent one $\rho_2(0)=|\psi(0)\rangle\langle\psi(0)|$ where $|\psi(0)\rangle=\left(|1\rangle+i\,|0\rangle\right)/\sqrt{2}$.

The quantification of the degree of non-Markovianity is also based on a metric which requires to contract under all completely positive trace-preserving maps. The trace distance might serve this aim, which is given as~\cite{Nielsen2000}:
\begin{equation}
    \Delta_{tr}(\rho_1||\rho_2)=\frac{1}{2}\mathrm{Tr}\left|\rho_1-\rho_2\right|,
\end{equation}
where $\left|A\right|=\sqrt{A^\dagger\,A}$. Based on the trace distance, the measure of the non-Markovianity is defined as \cite{Breuer2009, Laine2010}:
\begin{equation}
\label{eq:nonMar}
    \mathcal{N}=\mathrm{max}_{\rho_1,\rho_2}\int_{\sigma>0}\sigma(t)\,dt,
\end{equation}
where $\sigma(t)=\frac{d}{dt}\Delta_{tr}(\rho_1||\rho_2)$. The idea on such a quantification is based on the distinguishably of quantum states. A Markovian dynamics can always be represented by a dynamical semigroup which is completely positive and trace preserving. The trace distance always contracts under Markovian dynamics, showing the reduction in the distinguishably of two arbitrary quantum states. The flow of information is then considered outside of the system of interest. The rate of change of $\sigma(t)$ is negative, with the result that $\mathcal{N}$ in Eq.~(\ref{eq:nonMar}) is zero. On the other hand, non-Markovian dynamics can not be described by any completely positive and trace-preserving map. Under such a map, the distinguishably of two arbitrary states can increase for some time interval which indicates the flow of information back to the system. In this time domain, $\sigma(t)>0$. Please note that, the non-Markovianity measure in Eq.~(\ref{eq:nonMar}) accounts for the whole regions where $\sigma(t)>0$. Therefore, if $\mathcal{N}>0$, then the dynamics is always non-Markovian. In general, a numerical discretization method is employed overall pair of initial states to determine the non-Markovianity. Instead, in our calculations, choosing the orthogonal pair $\rho_1(0)$ and $\rho_3(0)=|0\rangle\langle0|$ would be sufficient to show the non-Markovian character that arises in our framework~\cite{Breuer2009,Laine2010}. 

\section{Results}
\label{sec:results}
Since we analyze the tunneling problem, there should always be a constant tunneling rate between two wells, that is, $\Delta_0=1$. The remaining parameters are free to choose which are scaled with respect to $\Delta_0$ and presented in dimensionless units ($\hbar=k_B=1$). In the following, we will investigate the barrier and its random disturbance on the coherence dynamics and non-Markovianity for the unbiased ($\epsilon=0$) and biased ($\epsilon=2.0$) cases. Since we neglect any quantum features of the environment leading to stochastic energy bias, we consider that it always exists but plays a secondary role, so we fixed $\kappa=0.1$ in our calculations.

\subsection{Static barrier}
\begin{figure}[hbt!]
\begin{center}
    \begin{tabular}[b]{c}
    \includegraphics[width=0.4\linewidth]{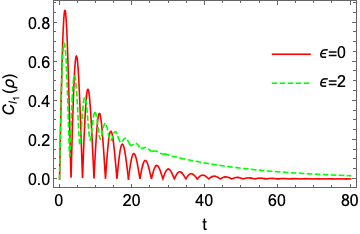}\\ \small (a) {Initial state: $\rho_1(0)$}
   \end{tabular}\qquad
    \begin{tabular}[b]{c}
   \includegraphics[width=0.4\linewidth]{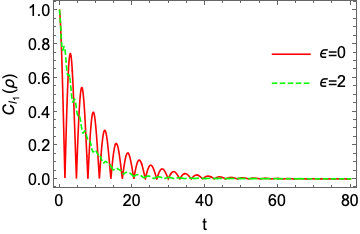}\\ \small (b) {Initial state: $\rho_2(0)$}
   \end{tabular}\qquad
    \end{center}
    \caption{$l_1$ norm of coherence $C_{l_1}(\rho)$ is plotted as a function of time $t$ for the static barrier case ($\Delta_1=0$) and the parameters $\Delta_0=1$, $\kappa=0.1$, $\epsilon=0$ (red, solid line) and $\epsilon=2.0$ (green, dashed line). (a) is plotted for the initially non-coherent state $\rho_1(0)$, while (b) is for the initially coherent one $\rho_2(0)$.}
\label{fig:cl1-versus-Delta1-static}
\end{figure}

In the first analyzed case, we consider that the barrier is static (that is, $\Delta_1=0$). The dynamical equations presented in Eq.~(\ref{eq:genEqu}) can be solved analytically. For the unbiased case ($\epsilon=0$) and the initial states considered $\rho_{1}(0)$ and $\rho_{2}(0)$, the $l_1$ norm of coherence takes simple forms,
\begin{eqnarray}
\label{eq:cl1_static}
C_{l_1}(\rho_1(0))&=&\left|\frac{e^{-\kappa\,t}\,\mathrm{sin}\left(\Omega\,t\right)}{\Omega}\right|,\nonumber\\
C_{l_1}(\rho_2(0))&=&\left|e^{-\kappa\,t}\frac{\Delta_0}{\Omega}\,\mathrm{cos}\left(t\,\Omega+\theta\right)\right|,
\end{eqnarray}
where $\Omega=\sqrt{\Delta_0^2-\kappa^2}$ and $\theta=\mathrm{arctan}\left(\kappa/\Omega\right)$. It is obvious from Eq.~(\ref{eq:cl1_static}) that $l_1$ norm of coherence exhibits damped oscillations. However, for the biased case, the analytical equations are too long and uninspiring. Therefore, in Fig.~\ref{fig:cl1-versus-Delta1-static}, we plot the dynamics of coherence for the biased and unbiased cases and the initial states $\rho_{1}(0)$ (Fig.~\ref{fig:cl1-versus-Delta1-static}(a)) and $\rho_{2}(0)$ (Fig.~\ref{fig:cl1-versus-Delta1-static}(b)). To elucidate the influence of the barrier on the coherence dynamics, we first consider the case where there is no barrier (that is, $\Delta_0=0$). In this scenario, there is only bias and its damping induced by its coupling to the high-temperature thermal environment. Therefore, $C_{l_{1}}(\rho)$ simply decays exponentially in time for the initial coherent state $\rho_2(0)$, while it always stays zero for the initial incoherent one $\rho_1(0)$ (see also Eq.~(\ref{eq:cl1_static})).  Introducing a barrier, on the other hand, substantially changes the coherence dynamics. As shown in Fig.~\ref{fig:cl1-versus-Delta1-static}(a), coherence is induced for the initial incoherent state $\rho_1(0)$. For both initial states, $C_{l_{1}}(\rho)$ shows damped oscillatory behavior. The damping is raised by its coupling to the thermal environment where the parameter $\kappa$ just controls how fast the coherence decays. In Fig.~\ref{fig:cl1-versus-Delta1-static}, unbiased ($\epsilon=0$) and biased ($\epsilon=2$) cases are also presented. It is clear from the figure that $\epsilon$ decreases the amplitude of oscillations. Note that, non-zero bias makes tunneling more difficult, since it introduces a gap between two localized potential wells~\cite{Iwans2000}. On the other hand, as shown by the comparison of Fig.~\ref{fig:cl1-versus-Delta1-static}(a) with Fig.~\ref{fig:cl1-versus-Delta1-static}(b), the coherence dynamics exhibits adverse behavior for the two initial states. Even the induced coherence (Fig.~\ref{fig:cl1-versus-Delta1-static}(a)) is more robust to dissipation for the biased case (green dashed line) compared to the unbiased one (red solid line), the behavior is reversed for the initial coherent state (Fig.~\ref{fig:cl1-versus-Delta1-static}(b)). 

We should emphasis that for the static barrier case, the Gaussian noise $\epsilon(t)$ modeled as the delta correlated function acts only as the source of the environment. Such noise can mimic the interaction of a system with quantum oscillators of a reservoir in the high temperature limit~\cite{Haken1973}. Under the approximation, the environment is purely classical, and there would be no back-flow of information from the environment to the system. After a straightforward calculation, the non-Markovianity measure in Eq.~(\ref{eq:nonMar}) for the initially orthogonal pair states $\rho_1(0)$ and $\rho_3(0)$ is found to be exactly zero. The result can also be interpreted from the master equation~(\ref{eq:masterEG}) which is obtained after averaging over Gaussian noise $\epsilon(t)$. The master equation is in the Lindblad form and can always be represented by a completely positive and trace preserving map~\cite{Laine2010}. The distinguishability is therefore reduced monotonically during the dynamical evolution. The only effect of $\kappa$ is to increase the dissipation which may reduce the amplitude of oscillations and to speed up the decay.

\subsection{Fluctuating barrier}
We next investigate the role of the barrier fluctuations on the quantum coherence and elucidate its non-Markovian character. There are two parameters that characterize the fluctuations in the barrier. One of them is $\Delta_1$ which gives the amplitude of the noise. The other one is $\nu$, which determines the fluctuation frequency. Instead of focusing on these parameters separately, we will consider the ratio known as Kubo number $K=\Delta_1/\nu$~\cite{Goychuk1995}. It is a proper measure for the color of the noise. For example, when the amplitude $\Delta_1$ is fixed, $K\ll1$ indicates a weakly colored noise due to the very fast barrier fluctuations, while $K\gg 1$ dictates strongly colored (slow) noise. In the case $K\ll1$, the system barely feels the effect of the noise. In this scenario, the barrier can be considered to be almost static and the results presented in the previous sections are the same. On the other hand, for the case where $K\gg 1$ the system dynamics is governed by the superposition of two solutions with tunneling rates $\Delta_{0}+\Delta_1$ and $\Delta_{0}-\Delta_1$. 

\begin{figure}[hbt!]
\begin{center}
    \begin{tabular}[b]{c}
    \includegraphics[width=0.4\linewidth]{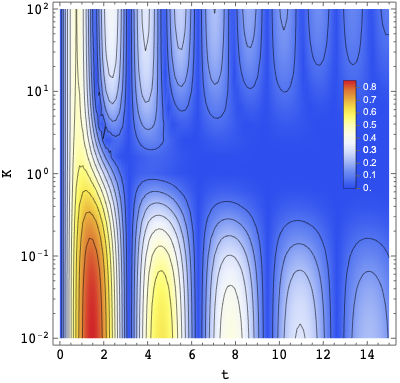}\\ \small (a) {$\epsilon=0$}
   \end{tabular}\qquad
    \begin{tabular}[b]{c}
   \includegraphics[width=0.4\linewidth]{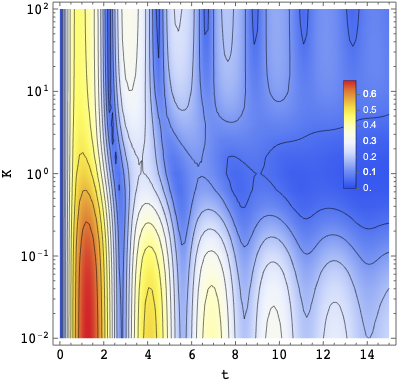}\\ \small (b) {$\epsilon=2$}
   \end{tabular}\qquad
    \begin{tabular}[b]{c}
    \includegraphics[width=0.4\linewidth]{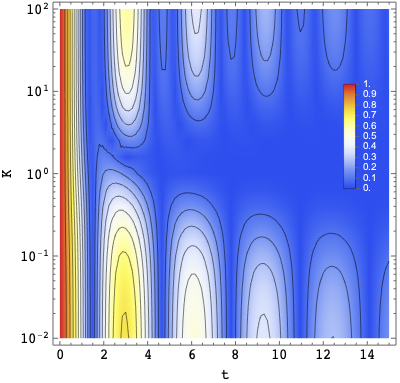}\\ \small (c) {$\epsilon=0$}
   \end{tabular}\qquad
    \begin{tabular}[b]{c}
   \includegraphics[width=0.4\linewidth]{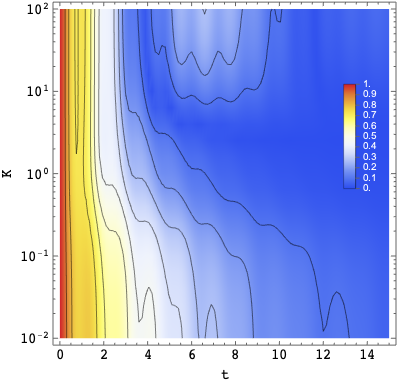}\\ \small (d) {$\epsilon=2$}
   \end{tabular}\qquad
    \end{center}
    \caption{Contour plot of $C_{l_1}(\rho)$ as a function of dimensionless time and Kubo number $K=\Delta_1/\nu$ for the parameters $\Delta_0=\Delta_1=1.0$, $\kappa=0.1$, $\epsilon=0$ ((a) and (c)) and  $\epsilon=2.0$ ((b) and (d)). While the sub-figures (a) and (b) are plotted for the initial incoherent state  $\rho_1(0)$,  (c) and (d) are for the initial coherent state $\rho_2(0)$. In the plots 10 equidistant contour lines are considered between zero and the maximum value.}
\label{fig:cl1-versus-Delta1-fluctuating}
\end{figure}

 For the fluctuating barrier case, it is not possible to obtain a simple analytical expression for the coherence measure. Therefore, we only analyze the results numerically. In Fig.~\ref{fig:cl1-versus-Delta1-fluctuating}, we present contour plots of the $l_1$ norm of coherence as a function of the Kubo number $K$ and the dimensionless time for the initial coherent and incoherent states. Both unbiased and biased cases are considered. The general observation from Fig.~\ref{fig:cl1-versus-Delta1-fluctuating} is that the  $C_{l_1}(\rho)$ displays a damped oscillating time dependence irrespective of the properties of the RTN noise. This finding can be accounted for by noting that both the thermal dephasing noise and RTN on the barrier height lead to the destruction of the coherence in the long time limit. For the initial incoherent state $\rho_1(0)$ (Fig.~\ref{fig:cl1-versus-Delta1-fluctuating}(a) and Fig.~\ref{fig:cl1-versus-Delta1-fluctuating}(b)), the maximum coherence is observed for the regions where Kubo number $K\ll 1$ corresponding to weakly colored noise. However, there is an exceptional region where $C_{l_1}(\rho)$ is independent of the value of $K$ in a very short time limit for the coherent initial state $\rho_2(0)$ (see Fig.~\ref{fig:cl1-versus-Delta1-fluctuating}(c) and Fig.~\ref{fig:cl1-versus-Delta1-fluctuating}(d)). It can also be seen from Fig.~\ref{fig:cl1-versus-Delta1-fluctuating} that the boundary $K\approx 1$ divides the dynamics of coherence into two different regimes with respect to the noise color, i.e., in the limit of $K\rightarrow 1$ (intermediate colored noise), coherence decays exponentially, while for the weak and strong colored noise, one can observe a strong death-rebirth cycle for coherence with the damping rate $\nu$ for all sub-figures of Fig.~\ref{fig:cl1-versus-Delta1-fluctuating}. Regardless of the value of $\epsilon$ and the initial state, the decoherence life time strongly depends on the Kubo number and has a resonance like behavior which decreases with increasing $K$ until the critical value $K=1$ and then tends to increase.

\begin{figure}[hbt!]
\begin{center}
    \begin{tabular}[b]{c}
    \includegraphics[width=0.4\linewidth]{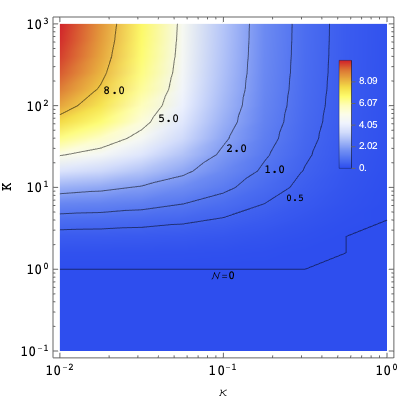}\\ \small (a) {$\epsilon=0$}
   \end{tabular}\qquad
    \begin{tabular}[b]{c}
   \includegraphics[width=0.4\linewidth]{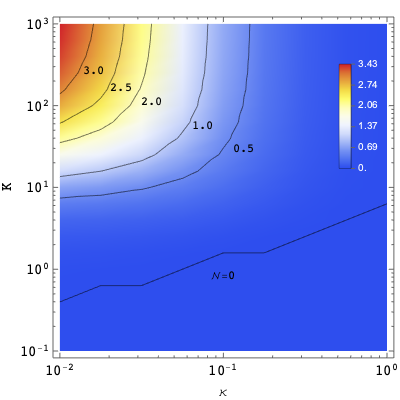}\\ \small (b) {$\epsilon=2$}
   \end{tabular}\qquad
   
    \end{center}
    \caption{Non-Markovianity $\mathcal{N}$ is plotted as functions of $\kappa$ and Kubo number $K=\Delta_1/\nu$ for the parameters $\Delta_0=\Delta_1=1.0$, $\epsilon=0$ (a) and $\epsilon=2.0$ (b). 6 contour lines are chosen between  minimum and maximum values of $\mathcal{N}$. Note that, $\mathcal{N}$ is computed for $\rho_1(0)$ and $\rho_3(0)$.}
\label{fig:nm-K-versus-kappa}
\end{figure}

 We finally discuss the non-Markovian feature of the random barrier fluctuations and its related effect on the quantum coherence. In this sense, we present some analytical results for the non-Markovianity measure (Eq.~(\ref{eq:nonMar})) when only RTN parameters are non-zero (i.e., $\epsilon=\kappa=\Delta_0=0$). Here, we are able to show the non-Markovian property of the barrier fluctuations. By solving the dynamical equations in Eq.~(\ref{eq:genEqu}) for the initial states $\rho_1(0)$ and $\rho_3(0)$, one can easily obtain the distinguishability as
 \begin{eqnarray}\label{eq:fbDist}
 \Delta_{tr}(\rho_1,\rho_3)=\left|\frac{e^{-\nu\,t}}{\gamma}\left(\gamma\,\mathrm{cos}(\gamma\,t)+\nu\,\mathrm{sin}(\gamma\,t)\right)\right|.
 \end{eqnarray}
 The trace distance exhibits damped oscillating behavior with noise frequency $\nu$ and $\gamma=-\sqrt{\Delta_1-\nu}$. Note that, when $\nu>\Delta_1$, $\gamma$ is purely imaginary, so $\Delta_{tr}(\rho_1,\rho_3)$ is a uniformly decreasing function of time, so $\mathcal{N}=0$. Using $\sigma(\rho_1,\rho_3)$ of Eq.~(\ref{eq:fbDist}) and the non-Markovianity in Eq.~(\ref{eq:nonMar}), $\mathcal{N}$ for $\Delta_1>\nu$ can be obtained analytically as
 \begin{equation}
 \mathcal{N}=\frac{1}{-1+e^{\frac{\pi}{\sqrt{K^2-1}}}}. 
 \end{equation}
 The obtained expression $\mathcal{N}$ strongly depends on the square of the noise color $K$. One can note that when $K\ll 1$ (almost static barrier), $\mathcal{N}=0$. For a more general case it is not possible to give an analytical expression for the non-Markovianity. Instead, for the general case, we display the results of numerical solutions in Fig.~\ref{fig:nm-K-versus-kappa} as the contour lines of $\mathcal{N}$ as the functions of the Kubo number $K$ and the thermal noise strength $\kappa$ for the system including both the thermal and external noise for the unbiased (Fig.~\ref{fig:nm-K-versus-kappa}(a)) and biased (Fig.~\ref{fig:nm-K-versus-kappa}(b)) cases. The most prominent observation from Fig.~\ref{fig:nm-K-versus-kappa} is that $\mathcal{N}$ increases with increasing $K$ and with decreasing $\kappa$ for unbiased and biased cases. Ref.~\cite{Cai2018} has also found that $\mathcal{N}$ increases with increasing Kubo number for non-equilibrium dephasing model. It is obvious from Fig.~\ref{fig:nm-K-versus-kappa} that the system dynamics is Markovian at high $\kappa$s. It can also be deduced from Fig.~\ref{fig:nm-K-versus-kappa}(a) (Fig.~\ref{fig:nm-K-versus-kappa}(b)) that Markovian-non-Markovian transition is (not) abrupt at $K=1$ line for unbiased (biased) case. This transition is also reported for a pure dephasing model with non-equilibrium noise by Cai and Zeng~\cite{Cai2016}. At the limit $K\rightarrow 0$, RTN approaches the white-noise~\cite{Goychuk1995}, which can create only Markovian dynamics. Therefore, as expected, we find that the dynamics is also Markovian for this parameter regime (see Fig.~\ref{fig:nm-K-versus-kappa}). On the other hand, the time dependence of coherence displays an opposite trend with the noise color. For small $K$, the decoherence time is longer for both initially coherent and incoherent states (see Fig.~\ref{fig:cl1-versus-Delta1-fluctuating}), which can be understood by examining the low and high frequency limits of RTN. 
 
 One of general insight from above findings is that the decoherence time does not show a monotonous trend with respect to Kubo number. That is, it decreases with increasing $K$ until $K=1$, and then starts to increase with increasing $K$ for initially coherent and non-coherent states in both unbiased and biased cases (see Fig.~\ref{fig:cl1-versus-Delta1-fluctuating}). $K=1$ seems to play an important role in both the existence of non-Markovianity and decoherence time. If one defines the decoherence rate is the inverse of decoherence time, then noise color dependence of decoherence rate displays a resonance structure with a peak at $K=1$. Also, at $K=1$ the coherence decays exponentially without any oscillations (see Fig.~\ref{fig:cl1-versus-Delta1-fluctuating}(a-d)). Hence, although there is no direct relation between coherence and non-Markovianity, both properties have special structures at around $K=1$.
 
 In some studies of coherence-Markovianity relation, authors report a direct connection between those two quantities for the model systems they investigate. For example, Ref.\cite{Addis2014} found that the stationary coherence is maximized for non-Markovian dynamics in an Ohmic spin-boson model at low temperature. As we study the effect of classical noise which is an infinite temperature limit, the stationary value of coherence is zero. But, Fig.~\ref{fig:cl1-versus-Delta1-fluctuating} and Fig.~\ref{fig:nm-K-versus-kappa} indicate that one can observe the decoherence rate is higher when the dynamics is non-Markovian, which seems to contradict the findings reported in Ref.~\cite{Addis2014, Bhattacharya2018, Cakmak2017,Radhakrishnan2019}. Ref.\cite{Bhattacharya2018} studied a global system–bath interaction in the both absence and presence of non-Markovian noise. Also, our findings are in contrast with those reported in  Ref.~\cite{Cakmak2017}  which studied a collisional model where both the system and environment are composed of
spin-$1/2$ particles and found that the higher non-Markovianity leads to low decoherence rate. On the other hand, Chanda and Bhattacharya~\cite{Chanda2016} who analyzed the dephasing channel, and found similar coherence-non-Markovianity relation to our results. Hence, our findings apparently indicates that the connection between non-Markovianity and decoherence is system-dependent and is not universal.

\section{Conclusions}
\label{sec:conc.}
We have studied the coherence and non-Markovianity of a quantum tunnelling system whose energy gap is subject to a delta-correlated Gaussian noise and tunneling rate is disturbed by a random telegraph noise. The main aim of the study was to check whether there exists any correlations between the coherence of the system and the non-Markovianity of the system dynamics. By averaging over the two separate noises, a master equation for the population, coherences and their RTN averages were obtained. It is found that, for a restricted range of system and noise parameters, analytical expressions for coherence measure-$C_{l_1}(\rho)$- and non-Markovianity-$\mathcal{N}$- derived for non-coherent and fully coherent initial states. The dynamics of the system is found to be non-Markovian only when the RTN barrier fluctuations satisfy $K\gg 1$, meaning that the noise amplitude is larger than the noise frequency. Contrary to the results of some similar studies on different toy models, we have found that one could not find a direct connection between the system coherence and non-Markovianity of the system dynamics except that both $\mathcal{N}$ and $C_{l_1}(\rho)$ change character around $K=1$. However, a consistent relation between the decoherence rate and non-Markovianity could be established for all the different initial states for both biased and non-biased two-level system. Surprisingly, we have found that decoherence rate of non-Markovian dynamics is higher compared to that of Markovian dynamics. As the non-Markovianity measure we have used is based on back-flow of information from the environment to the system, one would expect the coherence in such a scenario to decay slower which is reported for a number of systems~\cite{Addis2014, Bhattacharya2018, Cakmak2017,Radhakrishnan2019}. Since both coherence and non-Markovianity are considered as resources for quantum tasks, the current findings might be relevant in elucidating the relation between coherence of the system and non-Markovianity of the system dynamics.

\vspace{0.5in}
\textbf{Data availability} Data will be made available upon reasonable request.

\end{document}